\documentclass[twocolumn,prl,showpacs,amsfonts,amsmath,assymb,eufrak]{revtex4-1}
\usepackage{epsfig}
\usepackage{color}
\usepackage{bm}
\usepackage{graphicx}

\newcommand{\eq}[1]{\begin{equation} #1 \end{equation}}
\newcommand{\eqa}[2]{\begin{equation} #1 \label{#2} \end{equation}}
\newcommand{\balign}[1]{\begin{align} #1 \end{align}}

\newcommand{\figin}[4]
{\begin{figure}[tb]
\centering
\includegraphics[width= #1]{#2}
\caption{#3}
\label{f:#4}
\end{figure}}

\newcommand{\figinb}[4]
{
 \begin{figure}[b]\centering\includegraphics[width= #1]{#2}\caption{#3}\label{f:#4}\end{figure}
}

\newcommand{\todayd}{\the\year/\the\month/\the\day}
\newcommand{\del}{\partial}
\newcommand{\bib}{\bibitem}

\newcommand{\lb}{\label}
\newcommand{\nt}{\notag}

\newcommand{\bel}{\begin{easylist}}
\newcommand{\eel}{\end{easylist}}

\newcommand{\eref}[1]{Eq.~\eqref{#1}}

\newcommand{\fref}[1]{Fig.~\ref{f:#1}}

\def \({\left(}
\def \){\right)}
\def \[{\left[}
\def \]{\right]}
\newcommand{\la}{\left\langle}
\newcommand{\ra}{\right\rangle}

\newcommand{\sumtwo}[2]%
{\mathop{\sum_{#1}}_{#2}}
\newcommand{\sumthree}[3]%
{\mathop{\mathop{\sum_{#1}}_{#2}}_{#3}}
\newcommand{\sumfour}[4]%
{\mathop{\mathop{\mathop{\sum_{#1}}_{#2}}_{#3}}_{#4}} 
\newcommand{\prodtwo}[2]%
{\mathop{\prod_{#1}}_{#2}}
\newcommand{\mintwo}[2]%
{\mathop{\min_{#1}}_{#2}}
\newcommand{\maxtwo}[2]%
{\mathop{\max_{#1}}_{#2}}
\newcommand{\maxthree}[3]%
{\mathop{\mathop{\max_{#1}}_{#2}}_{#3}}
\newcommand{\limtwo}[2]%
{\mathop{\lim_{#1}}_{#2}}
\newcommand{\suptwo}[2]%
{\mathop{\sup_{#1}}_{#2}}
\newcommand{\supthree}[3]%
{\mathop{\mathop{\sup_{#1}}_{#2}}_{#3}}
\newcommand{\supfour}[4]%
{\mathop{\mathop{\mathop{\sup_{#1}}_{#2}}_{#3}}_{#4}} 
\newcommand{\inftwo}[2]%
{\mathop{\inf_{#1}}_{#2}}
\newcommand{\infthree}[3]%
{\mathop{\mathop{\inf_{#1}}_{#2}}_{#3}}
\newcommand{\inffour}[4]%
{\mathop{\mathop{\mathop{\inf_{#1}}_{#2}}_{#3}}_{#4}} 

\newcommand{\bsp}{\boldsymbol{p}}
\newcommand{\bsq}{\boldsymbol{q}}

\newcommand{\Di}{\mathit{\Delta}}

\newcommand{\para}[1]{{\em #1}\/.---}

\newcommand{\tlr}{\tilde{R}}
\newcommand{\pss}{p^{\rm ss}}
\newcommand{\peq}{p^{\rm eq}}
\newcommand{\dsgm}{\dot{\sigma}}
\newcommand{\dsgmhs}{\dot{\sigma}^{\rm HS}}

\def\rnum#1{\resizebox{0.5em}{\height}{\expandafter{\romannumeral #1}}}
\def\Rnum#1{\resizebox{0.5em}{\height}{\uppercase\expandafter{\romannumeral #1}}}

\makeatletter
\renewcommand{\@cite}[1]{\textsuperscript{#1)}}
\makeatother

\usepackage{graphicx}
\usepackage{dcolumn}
\usepackage{bm}
\usepackage{textcomp}
\usepackage{color,xcolor}

\begin{document}

\preprint{APS/123-QED}

\newcommand{\titlename}{Information-theoretical bound of the irreversibility in thermal relaxation processes}

\preprint{APS/123-QED}

\title{\titlename}

\author{Naoto Shiraishi}
\affiliation{Department of Physics, Gakushuin University, 1-5-1 Mejiro, Toshima-ku, Tokyo, Japan} 

\author{Keiji Saito}
\affiliation{Department of Physics, Keio university, Hiyoshi 3-14-1, Kohoku-ku, Yokohama, Japan}%

\date{\today}

\begin{abstract}
We establish that entropy production, which is crucial to the characterization of thermodynamic irreversibility, is obtained through a variational principle involving the Kulback-Leibler divergence. 
A simple application of this representation leads to an information-theoretical bound on entropy production in thermal relaxation processes; this is a stronger inequality than the conventional second law of thermodynamics. 
This bound is also interpreted as a constraint on the possible path of a thermal relaxation process in terms of information geometry. Our results reveal a hidden universal law inherent to general thermal relaxation processes.
\end{abstract}

\maketitle


\para{Introduction}
During the last two decades, significant achievements have been made in nonequilibrium statistical mechanics such as finding universal symmetry in fluctuating entropy production, which is manifested as the fluctuation theorem~\cite{ECM93, GC95, Kur98, Mae99} and the Jarzynski equality~\cite{Jar97}. This triggers various equalities~\cite{HS01,KPV07, SU12, Cel12, Ver14, SS15, SD16, NRJ17}, and the validity has been confirmed by elaborated experiments~\cite{FT-exp, Jar-exp, SU-exp}. The fluctuation theorem  remarkably reproduces the second law, i.e., the nonnegativity of the entropy production, and the fluctuation dissipation theorem and further universal information on higher order fluctuations~\cite{ES95, Gal96, SU08}, which are all universal relations, regardless of specific class of processes.

On the other hand, given a restricted class of irreversible processes, one can obtain more detailed properties of thermodynamic irreversibility that are stronger than the conventional second law. The examples include the recent thermodynamic uncertainty relation developed in the class of the steady state systems with a finite current such as heat and charge currents. 
In this class, a significant constraint between the current fluctuation and the average current has been elucidated~\cite{BS15, Ging16, GRH17, BHS18, DS18}. Another example is the class of cyclic heat engines controlled with a finite speed, where the trade-off relation between thermodynamic efficiency and finite power is derived ~\cite{BSS15, SST16, ST17}. Such strong relations have helped us to deepen our understanding of the thermodynamic irreversibility inherent to each class of nonequilibrium process.

In this Letter, we further continue towards this direction by considering the class of {\it thermal relaxation processes}, which has not been sufficiently explored so far. We present several characteristics of thermodynamic irreversibility inherent to this class. Thermal relaxation phenomena are ubiquitous in nature and are one of the most critical targets in nonequilibrium physics. In addition, they are diverse and can be highly nontrivial even within Markovian dynamics. Metastable potentials are well known to induce various phenomena such as the Griffith phase in magnetic alloys~\cite{Bra87}, nonmonotonic relaxations~\cite{LR17}, and slow relaxations in glassy systems~\cite{TBF04}. Provided that a strong constraint on irreversibility exists in relaxation processes, this constraint can be salient information on many nontrivial phenomena because such a constraint inevitably leads to a general limitation on the possible relaxation path.

Our strategy in this regard is to use the information-theoretical argument. Here, special attention is paid to two important quantities in the relaxation processes: One is thermodynamic entropy production and the other is the Kullback-Leibler divergence (relative entropy) in information theory, which quantifies how far a given two probability distributions are. 
We first consider the entropy production for a general class of stochastic dynamics, which can contain time-dependent parameters. Then, we develop a variational principle for entropy production using the Kullback-Leibler divergence with two distributions that respectively evolve forward and backward in time  (shown in \eref{sgm-max}). This principle is convenient to consider the thermodynamic irreversibility in thermal relaxation processes where no time-dependent control parameters are involved. Based on the variational principle, we obtain a lower bound for entropy production in a relaxation process with the Kullback-Leibler divergence between the initial and the final distributions (shown in \eqref{relax}). Further, this relation leads to a characterization of the possible time evolution from the viewpoint of information geometry. We also show that the variational principle is connected to the fluctuation-theorem-type equality.

\para{Setup and variational expression of entropy production rate}
We start with the core part, namely, the variational-principle expression for the entropy production rate. We follow the standard framework of stochastic thermodynamics. Consider a system with discrete states attached to a heat bath with inverse temperature $\beta$. This heat bath induces stochastic transitions between different states. The probability distribution of state $i$ at time $t$, denoted by $p_i(t)$, follows
\eq{
\frac{d}{dt}p_i (t) =\sum_j R_{ij} (t) p_j (t),  \label{dynamics}
}
where $R_{ij} (t)$ is the transition matrix element from the state $j$ to $i$, which can be time-dependent. 
Throughout this Letter, we assume the detailed-balance condition 
$R_{ij}(t)e^{-\beta E_j(t)}=R_{ji}(t)e^{-\beta E_i(t)} \, , 
$
where $E_i (t)$ is the instantaneous energy of state $i$ at time $t$ and we set the Boltzmann constant to unity. The detailed-balance condition embodies the microscopic time reversibility in the equilibrium state. In the supplementary material, we also discuss the case without the detailed-balance condition and we present a similar result to the main result given below \cite{suppl}. We remark that a dynamics with continuous variables can be properly analyzed by using the discretized representation~\cite{SST16}, and therefore, one can safely continue the argument with the discrete picture.

\figin{7.0cm}{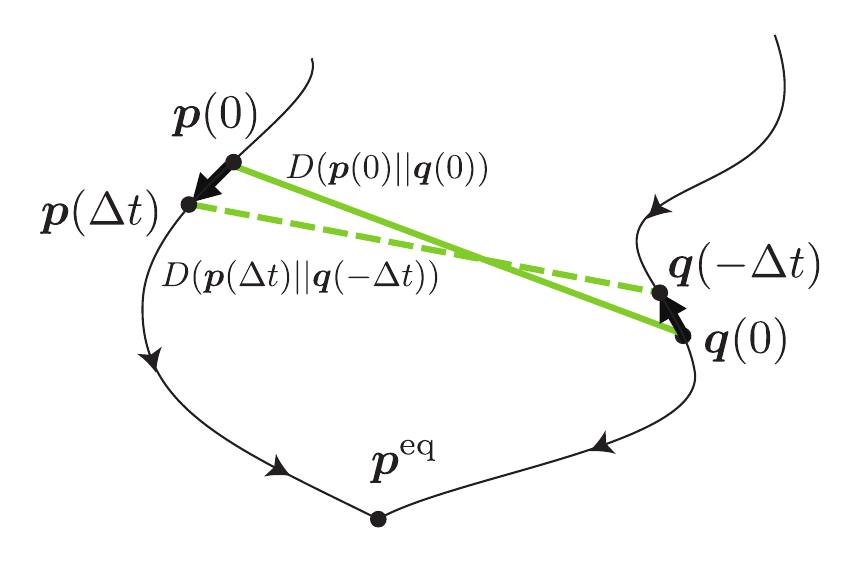}{
We draw a state space of probability distributions.
Two thin curves represent trajectories of probability distributions in dynamics with the transition matrix $R$.
Green lines represent the Kullback-Leibler divergence between two distributions.
The variational principle \eqref{sgm-max} claims that the difference between two Kullback-Leibler divergences $D(\bsp(0)||\bsq(0))-D(\bsp(\Di t)||\bsq(-\Di t))$ (solid line minus dashed line) is maximized when $\bsq(0)=\bsq(-\Di t)=\bsp^{\rm eq}$.
}{max-image}

The total entropy production from $t=0$ to $t=\tau$ is given by the integration of the entropy production rate
$\sigma_{[0,\tau]}  =  \int_0^{\tau} dt \,  \dsgm (t)$,
where the entropy production rate $\dot{\sigma}(t)$ at time $t$ is obtained as the sum of two contributions from the system and the heat bath~\cite{Sei12}:
\begin{align}
\dsgm (t) &:= \dsgm_{\rm sys} (t)+ \dsgm_{\rm bath} (t)\, ,  \nonumber \\
 \dsgm_{\rm sys} (t) &= \frac{d}{dt}\sum_i \( -p_i \ln p_i\) \, ,  \nt \\
   \dsgm_{\rm bath} (t) &=-\sum_{i}\beta E_i (t) \dot{p}_i \, .  \nonumber 
\end{align}
The system's entropy $\sigma_{\rm sys}$ is defined by the Shannon entropy. The entropy increase in the heat bath $\sigma_{\rm bath}$ is given through the amount of heat flow into the bath as in the conventional thermodynamics.

We here state the variational expression of the entropy production rate, which is our first main result in this Letter:
\eqa{
\dsgm (t) =\max_q \[ -\frac{d}{dt}D({\bm p}(t)||{\bm q}(-t))\] ,
}{sgm-max}
where $D$ is the Kullback-Leibler divergence~\cite{CTbook} (relative entropy) defined as $D({\bm p}||{\bm p}'):=\sum_i p_i \ln (p_i/p'_i)$.
The distribution ${\bm q}(-t)$ is the probability distribution that evolves backward in time with the same dynamics, which is introduced independently of the target distribution ${\bm p}$. In other words, the equation of motion is given by $-(d/dt)q_i(-t)=\sum_j R_{ij}(t) q_j (-t)$ (see also \fref{max-image}). 
We provide the proof at the end of this Letter, and here we focus on physics given by this relation.

The entropy production rate is obtained as the maximizing problem of the right-hand side in (\ref{sgm-max}) over all possible probability distributions ${\bm q}$. The maximum is achieved when ${\bm q}(-t)$ is the equilibrium distribution with instantaneous energies $\bsp^{\rm eq}$, i.e., $\dsgm=-\frac{d}{dt} D(\bsp (t)||\bsp^{\rm eq})$. Otherwise, the function defined in curly brackets is smaller than the entropy production rate, i.e., $\dsgm\geq -\frac{d}{dt}D({\bm p}(t)||{\bm q}(-t))$ for any $\bsq$. The right hand side in this inequality gives a lower bound on the entropy production, and hence the positive value is required to obtain thermodynamically nontrivial relation. As we show below, appropriate choice of $\bsq$ in specific class of dynamics makes the bound positive, which leads to a stronger bound on the thermodynamic irreversibility. The most important application of the relation (\ref{sgm-max}) in the context of the thermodynamic irreversibility is its application to the thermal relaxation processes as demonstrated below. In the supplementary material~\cite{suppl}, we also present another application to time-dependent processes with a systematic protocol which characterizes the thermodynamic irreversibility.

\para{Bound for relaxation processes}
We here show a simple but important application of the variational principle to thermal relaxation processes, that is, a dynamics with a {\em time-independent} transition matrix. We now derive an information-theoretical bound for the entropy production in this process. We set ${\bm q}(0)={\bm p}(\tau)$ in the relation (\ref{sgm-max}) and integrate both sides from initial time $t=0$ to $\tau/2$ to obtain the inequality $\sigma_{[0,\tau/2]}\geq D({\bm p}(0)||{\bm p}(\tau))$. From the monotonic increase of entropy production in time, one can immediately obtain the inequality on entropy production: 
\eqa{
\sigma_{[0,\tau]}\geq D({\bm p}(0)||{\bm p}(\tau)) \, .
}{relax}
Most crucial aspect of this relation is that it imposes a stronger constraint on the entropy production than the standard second law of thermodynamics, since the Kullback-Leibler divergence is always nonnegative. This result is valid for arbitrary dynamics as long as the transition matrix is time-independent. This strong constraint on the entropy production (\ref{relax}) is our second main result in this Letter.

At the initial time, the Kullback-Leibler divergence in the right-hand side is zero, and therefore, the positivity of the entropy production rate guarantees the above inequality in the early time stage~\cite{fn-early}. However, we stress that this inequality holds for the whole time region. Equality is achieved at both $\tau =0$ and $\tau \to \infty$. In the latter case, the distribution is the equilibrium distribution.

\figin{8.5cm}{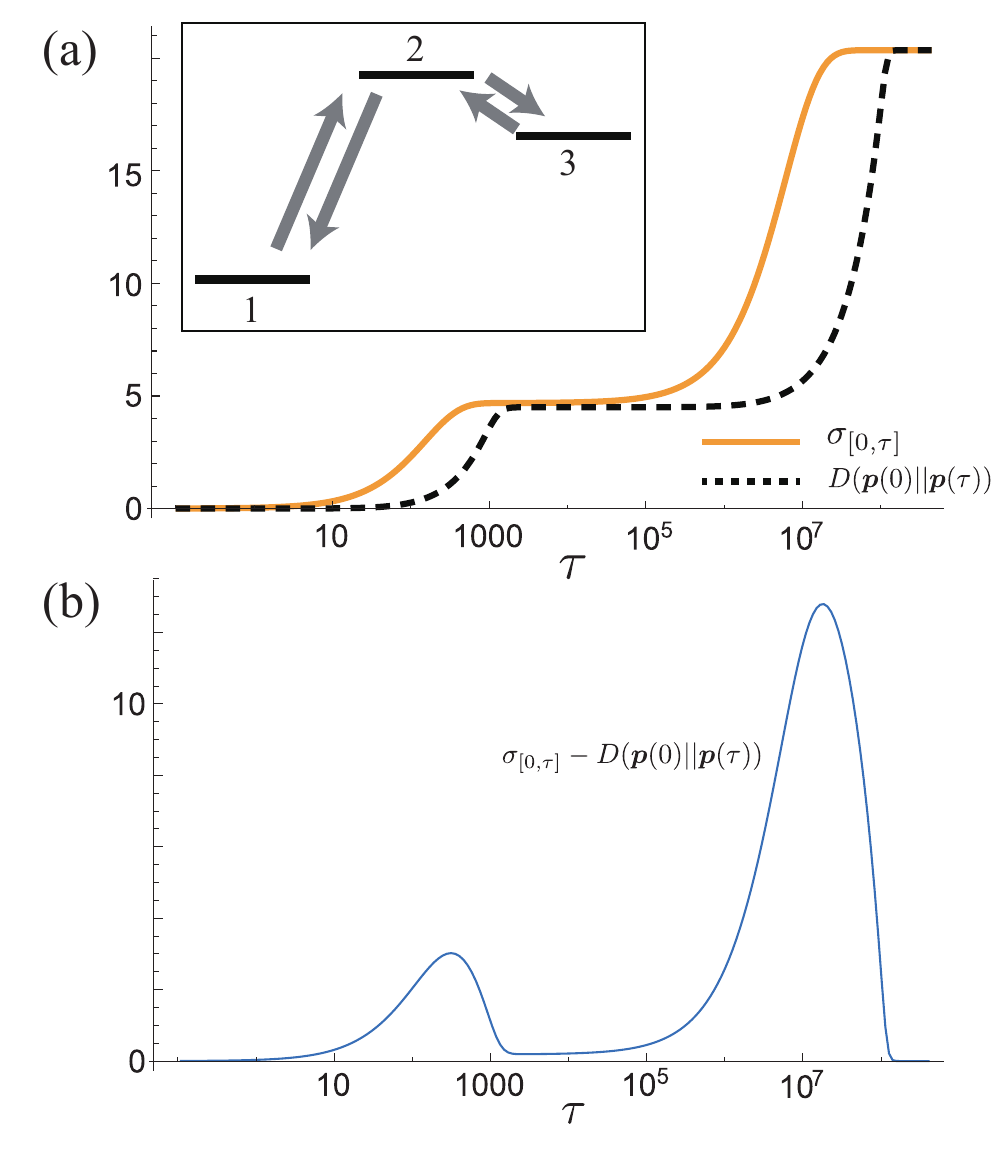}{
Demonstration of the relation (\ref{relax}) in a three-state toy model. The dynamics shows a long-time trap dominated by the states $2$ and $3$. (a) We plot $\sigma_{[0,\tau]}$ (orange solid line) and $D({\bm p}(0)||{\bm p}(\tau))$ (black dashed line). Although the entropy production shows anomalous two-step relaxation, $D({\bm p}(0)||{\bm p}(\tau))$ robustly bounds the entropy production. (b) We plot the difference between these two.}{plot}

We present an example with a three-state model to show how the inequality works for the whole time region. We set the energies of the three states as $(\beta E_1, \beta E_2, \beta E_3)=(6,30,24)$. See the inset of \fref{plot}.(a) for the energy structure. The transition rate is given in the form $R_{ij}=\gamma_{ij} e^{\beta(E_i-E_j)/2}/\left[ 2 \cosh( \beta(E_i-E_j)  ) \right] $ with the parameters $(\gamma_{12},\gamma_{23},\gamma_{31})=(10,0.1,0)$ and $\gamma_{ij}=\gamma_{ji}$. For these parameters, the transition rates are $R_{32}\sim 4.9 \times 10^{-3}, R_{12}\sim 6.1 \times 10^{-5}, R_{23}\sim 1.2 \times 10^{-5}$, $R_{21}\sim 2.1\times 10^{-15}$, and $R_{13}=R_{31}=0$. The initial distribution is set as $(p_1,p_2,p_3)=(0.1,0.8,0.1)$.
These transition rates indicate that the dynamics is dominated by the relaxation from the state $2$ to $3$ at the initial stage, and after this process the complete equilibration is achieved with the extremely long time. The plot in \fref{plot} shows nonmonotonic thermal relaxation and an exponentially long equilibration time, which support the scenario described here.
Most importantly, the figure clearly shows that the inequality \eqref{relax} robustly holds at all times. Intriguingly, a significant decrease in the difference between the entropy production and the Kullback-Leibler divergence is observed at the intermediate time~\cite{fn-sepa}. Such a reduction can be seen generically for systems with local equilibration~\cite{suppl}.

We remark that the relation \eqref{relax} applies not only to discrete systems but also to spatially continuous systems. In the supplementary material, we demonstrate an example where a Brownian particle trapped by a harmonic potential in one dimension \cite{suppl}. The validity of \eqref{relax} is analytically confirmed in this model.

\figin{7.5cm}{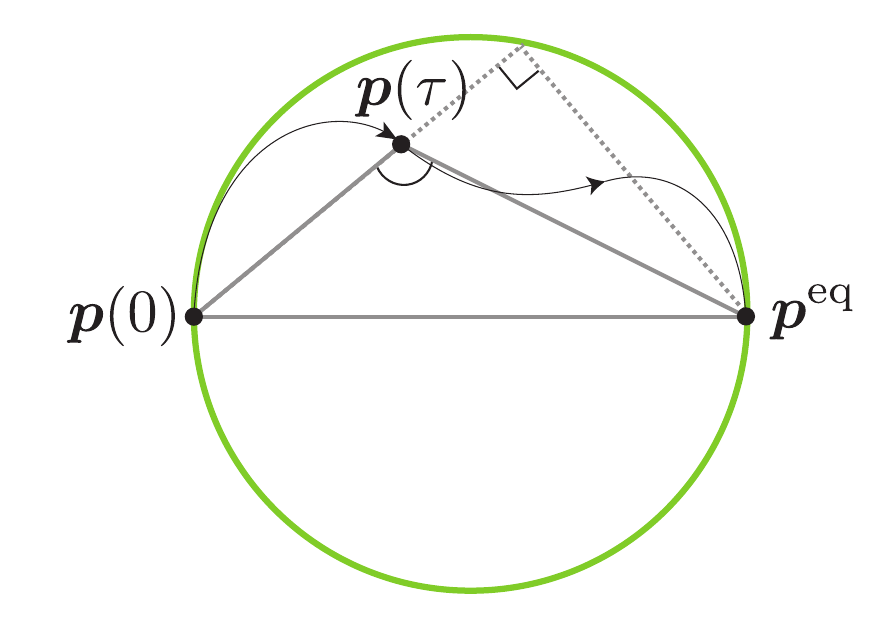}{
  Schematic of the constraint \eqref{infogeo} in terms of the information geometry. As a natural extension of the Pythagorean theorem in the information geometry, the inequality is interpreted as that the probability distribution at $\tau$ forms an obtuse angle and its path is inside the green ``circle". We emphasize that these relations are stronger than the conventional second law: $\sigma_{[0,\tau]} =D({\bm p}(0)||{\bm p}^{\rm eq}) - D({\bm p}(\tau )|| {\bm p}^{\rm eq}) \ge 0$.}{relax}

\para{Several interpretations: the information geometry and the speed limit}
Let us combine the standard expression of entropy production $  \sigma_{[0,\tau]}  =D({\bm p}(0)||{\bm p}^{\rm eq}) - D({\bm p}(\tau )|| {\bm p}^{\rm eq}) $ and the relation (\ref{relax}). Then we obtain a nontrivial constraint on the thermal relaxation process in terms of the information geometry:
\eqa{
  D({\bm p}(0)||{\bm p}^{\rm eq}) \geq D({\bm p}(0)||{\bm p}(\tau))  + D({\bm p}(\tau )|| {\bm p}^{\rm eq}) \, .
}{infogeo}
In the field of information geometry~\cite{ANbook}, the equality satisfied between three Kullback-Leibler divergences is called the Pythagorean theorem by regarding each contribution of the Kullback-Leibler divergence as a squared value of the distance forming a right-angled triangle. In this context, the inequality (\ref{infogeo}) implies that the probability distribution at an arbitrary time should be located at the point forming the obtuse angle. Thus the relaxation path must be confined inside the {\it circle} with a diameter determined by $\bsp(0)$ and $\bsp^{\rm eq}$, as shown schematically in \fref{relax}. The right angle is achieved at $\tau=0$ and $\tau=\infty$ where the system relaxes to the thermal equilibrium. It is remarkable that the universal constraint exists in the information-geometrical context regardless of each detail of the phenomenon.

We also mention that the relation (\ref{relax}) can be interpreted in terms of the speed limit in the stochastic thermodynamics \cite{SFS18}. We can rewrite this relation as $\tau \ge D({\bm p}(0)||{\bm p}(\tau)) / \bar{\sigma}$, where $\bar{\sigma}$ is the average entropy production rate $\bar{\sigma}:=\sigma_{[0,\tau]}/\tau$. This inequality is interpreted as a bound of the relaxation time. At the qualitative level, this expression implies that for a given distance between a target (final) distribution and an initial distribution, a large entropy production rate is necessary for a short relaxation time. In Ref.~\cite{SFS18}, the speed limit expression in the stochastic processes is discussed in a general class of Markovian dynamics, and it is shown that the dynamical activity as well as the average entropy production rate is crucial to describe the bound of manipulation time. However the present case contains only the entropy production rate, which is one of characteristics inherent to the class of thermal relaxation processes.

\para{The fluctuation-theorem-type equality}  
As is well known, the second law can be obtained from the fluctuation theorem. We consider whether the relation (\ref{sgm-max}) can be obtained through the fluctuation theorem. We remark that the thermodynamic uncertainty relation are considered not to be obtained from a fluctuation-theorem-like equality~\cite{Shi17-note, PB17}. Here, we show that in the present case, there exists a fluctuation-theorem-type equality that leads to the inequality \eqref{sgm-max}, while the original fluctuation theorem is not available.  

Let us consider a process with {\it time-independent} transition rates.
We integrate the expression \eqref{sgm-max} for a finite time $\Delta t$ to get
\eqa{
\sigma_{[0,\Di t]}\geq D(\bsp (\Di t)||\bsq(0))-D(\bsp(0)||\bsq(\Di t)),
}{sgm-max-disc}
where both $\bsp$ and $\bsq$ evolve in time with the same transition rates.
We present the fluctuation-theorem-type equality that reproduces this expression. Let $\Gamma_{i\to j}$ be a trajectory of the states during the time interval $\Di t$ from the initial state $i$ to the final state $j$. Then, one can derive the following equality:
\eqa{
\la \exp \[ -\hat{\sigma}(\Gamma_{i\to j})-\ln \frac{p_j(\Di t)}{q_j(0)}+\ln \frac{p_i(0)}{q_i(\Di t)}\] \ra =1,
}{FT-type}
where $\la ... \ra$ is the average over all possible paths and the initial distribution, and $\hat{\sigma}(\Gamma_{i\to j})$ is the stochastic entropy production. Details on the derivation of this equality and an extension to the case of time-dependent transition matrix are presented in the supplementary material~\cite{suppl}. One can readily check that \eref{sgm-max-disc} is obtained through this equality using Jensen's inequality and the relation $\sum_i \int d\Gamma_{i\to j}p_i(0)P(\Gamma_{i\to j})=p_j(\Di t)$.

\para{Proof of \eref{sgm-max}}
Here, we present a proof of \eqref{sgm-max}. We prove the following inequality that is equivalent to \eqref{sgm-max}: 
\eqa{
\frac{d}{dt}\[ D({\bm p}(t)||{\bm q}(-t))-D({\bm p}(t)|| {\bm p}^{\rm eq}) \] \geq 0 \, .
}{sgm-max-trans}
Note that the time-derivative does not act on the instantaneous equilibrium distribution ${\bm p}^{\rm eq}$.
The left-hand side is calculated as follows.
\begin{align}
  &\frac{d}{dt}\[ D({\bm p}(t)||{\bm q}(-t))-D({\bm p}(t)||{\bm p}^{\rm eq} ) \] \nt \\
=& \frac{d}{dt}\[ \sum_i p_i(t) \ln \( \frac{\peq _i}{q_i(-t)}\) \] \nt \\
  =& \sum_{i\neq j} R_{ij}p_j \ln \( \frac{\peq _i q_j}{\peq_j q_i}\) +\sum_{i\neq j} p_i \frac{R_{ij} q_j}{q_i} +\sum_i R_{ii}p_i \, .
     \end{align}
From the second line, we dropped $t$ dependence in the variables for the sake of simplicity.
Exchanging the indices $i$ and $j$ and using the detailed-balance condition $R_{ij}\peq_j=R_{ji}\peq_i$, we obtain the desired result as
\begin{align}
(8)=& \sum_{i\neq j} R_{ij}p_j \ln \( \frac{R_{ij} q_j}{R_{ji} q_i}\) +\sum_{i\neq j} R_{ij} p_j \frac{R_{ji} q_i}{R_{ij}q_j} -\sum_{i\neq j} R_{ij}p_j \nt \\
=& \sum_{i\neq j} R_{ij}p_j \[ \frac{R_{ji} q_i}{R_{ij}q_j} -1- \ln \( \frac{R_{ji} q_i}{R_{ij} q_j}\) \] \nt \\
\geq & 0.
\end{align}
In the last line, we used $x-1-\ln x\geq 0$ for any $x>0$.

\para{Discussion}
In this Letter, we present the information-theoretical argument in the thermal relaxation processes. The variational principle is a core to develop this theory, which demonstrates the deep connection between nonequilibrium statistical physics and information theory. 
The significance of the inequality (\ref{relax}) is that the amount of entropy production is bounded from below by the information-theoretical entropy determined only by the initial and final distributions. This is the remarkable constraint that is stronger than the conventional second law, which is imposed on the class of thermal relaxation processes. We remark that the variational principle \eqref{sgm-max} is applicable to general processes, and hence we hope that it helps in developing the other useful thermodynamic relations.

The inequality \eqref{relax} can be interpreted as the information-geometrical constraint (\ref{infogeo}). This provides a new approach to characterize the thermal relaxation processes. Further analysis with the information geometry techniques~\cite{ANbook,SB83,And88, Rup95,Cro07,Ito18} may help in finding richer properties of thermal relaxation. 

\para{Acknowledgement}
We express special thanks to Makiko Sasada for giving us a hint to arrive at final results. 
She has already derived a very similar inequality to (3).  
We are grateful to Shin-ichi Sasa and Hal Tasaki for fruitful discussions.
NS was supported by JSPS Grants-in-Aid for Scientific Research Grant Number JP19K14615. 
KS was supported by JSPS Grants-in-Aid for Scientific Research (JP16H02211 and JP17K05587).

\clearpage

\pagestyle{empty}

\makeatletter
\long\def\@makecaption#1#2{{
\advance\leftskip1cm
\advance\rightskip1cm
\vskip\abovecaptionskip
\sbox\@tempboxa{#1: #2}%
\ifdim \wd\@tempboxa >\hsize
 #1: #2\par
\else
\global \@minipagefalse
\hb@xt@\hsize{\hfil\box\@tempboxa\hfil}%
\fi
\vskip\belowcaptionskip}}
\makeatother
\newcommand{\vo}{\upsilon}
\newcommand{\midskip}{\vspace{3pt}}

\setcounter{equation}{0}
\def\theequation{A.\arabic{equation}}

\begin{widetext}

\begin{center}
{\large \bf Supplementary Material for  \protect \\ 
  ``Information-theoretical bound of the irreversibility in thermal relaxation processes'' }\\
\vspace*{0.3cm}
Naoto Shiraishi$^{1}$ and Keiji Saito$^{2}$ \\
\vspace*{0.1cm}
$^{1}${\small \it Department of Physics, Gakushuin University, 1-5-1 Mejiro, Toshima-ku, Tokyo, Japan} \\
$^{2}${\small \it Department of Physics, Keio University, Yokohama 223-8522, Japan} 
\end{center}

\setcounter{equation}{0}
\renewcommand{\theequation}{S.\arabic{equation}}

\section{Another application of the relation (2): irreversibility in forward and backward process}

In this section, we provide another application of \eref{sgm-max} in the main text.
In this application, we consider a systematic protocol which reveals the thermodynamic irreversibility for an arbitrary time-dependent processes for $0\leq t\leq \tau$.

Consider a time-dependent process in $0\leq t\leq \tau$. Let ${\bm R}(t)$ ($0\leq t\leq \tau$) be the transition matrix of the process.
We introduce its time-reversal operation given by ${\bm R}^\dagger (t):={\bm R}(2\tau -t)$ ($\tau\leq t\leq 2\tau$) and apply it immediately after the process driven by the transition matrix ${\bm R}(t)$ for $0\leq t\leq \tau$.
Note that after this time-reversal operation, i.e., at $t=2 \tau$, all the control parameters of the system return to the original values.
We denote by ${\bm p}$ the initial probability distribution at $t=0$, and by ${\bm p}'$ the final distribution at $t=2\tau$.
If the process is quasistatic, then ${\bm p}'={\bm p}$ holds, and otherwise ${\bm p}'$ differs from ${\bm p}$.
We employ the Kullback-Leibler divergence from ${\bm p}'$ to ${\bm p}$, $D({\bm p}||{\bm p}')$, as the measure of difference between two states~(\fref{irre}).

From \eref{sgm-max} in the main text, this protocol yields the following clear relation between these two quantities:
\eqa{
\sigma_{[0,\tau]}\geq D({\bm p}||{\bm p}')  .
}{irre-bound}
To get this inequality, we set ${\bm q}(0)={\bm p}'$ and integrate the both sides in the relation (2) over $0\leq t\leq \tau$. The left-hand side concerns only the forward process, while we need the backward process to obtain the right-hand side.
Notably, this inequality is a stronger bound than the conventional second law of thermodynamics.

\figinb{8cm}{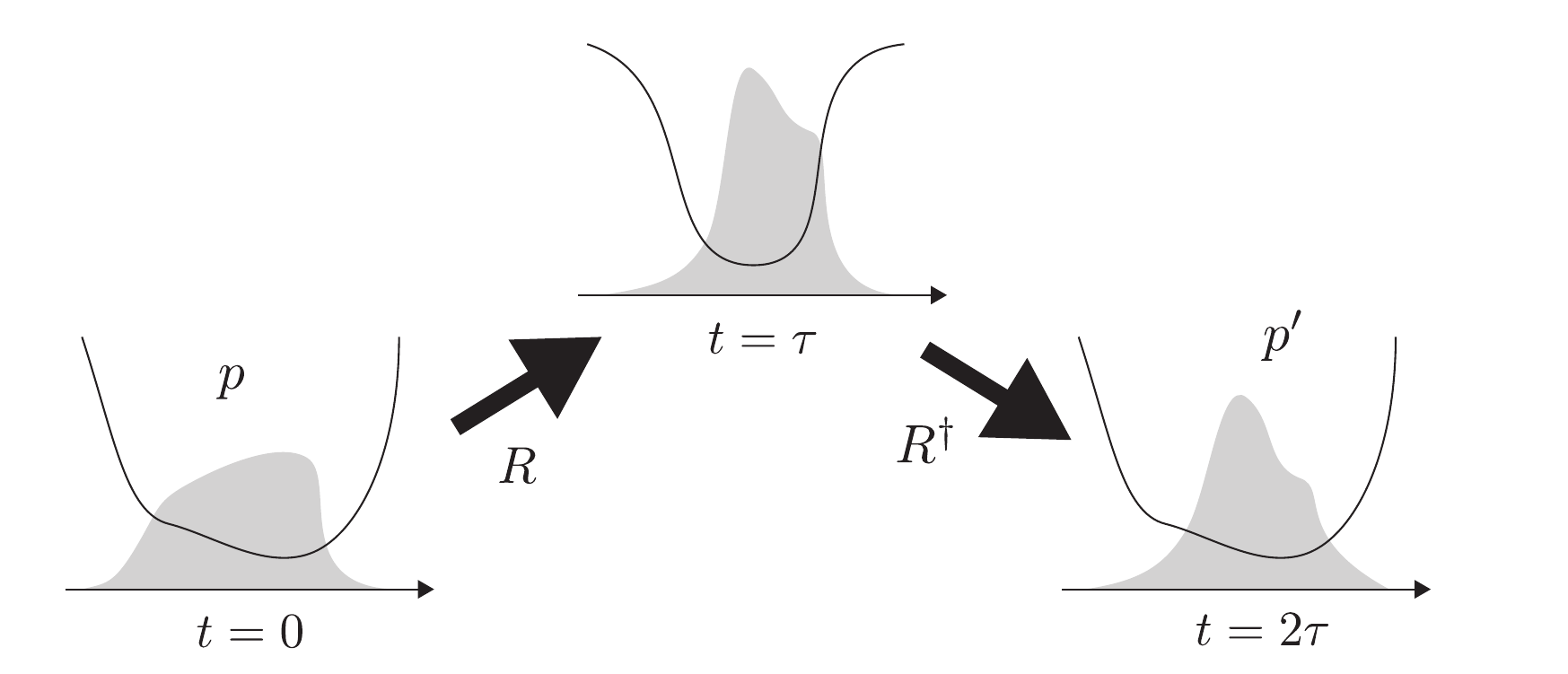}{We apply a forward process $R$ and its backward process $R^\dagger$ subsequently. The distribution function evolves in time with the transition matrix ${\bm R}$ for $0\leq t\leq \tau$, and ${\bm R}^\dagger$ for $\tau\leq t\leq 2\tau$. The distributions at $t=0$ and $t=2\tau$ are denoted by ${\bm p}$ and ${\bm p}'$, respectively. If ${\bm R}$ gives a quasistatic process, then $D({\bm p}||{\bm p}')=0$, while otherwise $D({\bm p}||{\bm p}')\geq 0$. We use $D({\bm p}||{\bm p}')$ as the measure of the irreversibility.}{irre}

\section{Generalization of the relation (2)}
Consider a general stochastic process with a transition matrix ${\bm R}(t) $ which does not satisfy the detailed-balance condition in general.
Let $\tlr_{ij} (t) :=R_{ji} (t) \pss_i  /\pss_j $ be the dual matrix of the transition matrix ${\bm R}(t)$, and ${\bm q}(t)$ evolves with the dual transition matrix $\tilde{\bm R}$. Here, ${\bm p}^{\rm ss}$ is the instantaneous steady state distribution.
Note that ${\bm R}$ and $\tilde{\bm R}$ has the same instantaneous steady state distribution. Then, for any concave function $f(x)$, we claim the following inequality for the $f$-divergence like quantity
\eqa{
\frac{d}{dt}\[ \sum_i p_i(t) f\( \frac{q_i(-t)}{\pss_i }\) \] \leq 0,
}{f-max}
which is the generalization of \eref{sgm-max}.
Note that the time-derivative does not act on the steady state distribution.

Before proving (\ref{f-max}), we demonstrate that the relation (\ref{f-max}) contains \eref{sgm-max} as its special case.
By setting $f(x)=\ln x$, the left-hand side of (\ref{f-max}) is transformed into
\eq{
\frac{d}{dt}\[ \sum_i p_i(t) \ln \( \frac{q_i(-t)}{\pss_i}\) \] = \frac{d}{dt}\[ D(\bsp(t)||{\bm p}^{\rm ss})-D(\bsp(t)||\bsq(-t)) \] = -\frac{d}{dt}D(\bsp(t)||\bsq(-t)) -\dsgmhs (t) ,
}
where 
\eq{
\dsgmhs (t) :=\sum_{i,j}R_{ij} (t) p_j (t) \ln \frac{\pss_i (t) p_j (t)}{\pss_j (t)  p_i (t)}
}
is the Hatano-Sasa entropy production rate~\cite{HS01}.
Since the substitution of ${\bm q} (-t)={\bm p}^{\rm ss}$ leads to the equality of the inequality (\ref{f-max}), we arrive at
\eq{
\dsgmhs (t) =\max_{q} \[ - \frac{d}{dt}D(p_i(t)||q_i(-t)) \] .
}
For systems with the detailed-balance condition, the Hatano-Sasa entropy production is equal to the thermodynamic entropy production and the above expression reduces to \eqref{sgm-max}.

We now prove the inequality (\ref{f-max}).
We transform the right-hand side as
\balign{
\frac{d}{dt}\[ \sum_i p_i(t) f\( \frac{q_i(-t)}{\pss_i}\) \] 
&=\sum_{i,j} \[ R_{ij}p_j f\( \frac{q_i}{\pss_i}\) -p_j \tlr_{ji}\frac{q_i}{\pss_j}f'\( \frac{q_j}{\pss_j}\) \] \nt \\
&=\sum_j p_j \sum_{i(\neq j)}\[ R_{ij} f\( \frac{q_i}{\pss_i}\) -\tlr_{ji}\frac{q_i}{\pss_j}f'\( \frac{q_j}{\pss_j}\) -R_{ij}f\( \frac{q_j}{\pss_j}\) +R_{ij}\frac{q_j}{\pss_j} f'\( \frac{q_j}{\pss_j}\) \] \nt \\
&=-\sum_j p_j \sum_{i(\neq j)}R_{ij}  \[ f\( \frac{q_j}{\pss_j}\) - f\( \frac{q_i}{\pss_i}\) +\frac{\tlr_{ji}}{R_{ij}}\frac{q_i}{\pss_j}f'\( \frac{q_j}{\pss_j}\) -\frac{q_j}{\pss_j} f'\( \frac{q_j}{\pss_j}\) \] , \lb{gen-mid}
}
where in the first line we exchange the indices $i$ and $j$.
By substituting $\tlr_{ji}/R_{ij}=\pss_j/\pss_i$ and denoting $x=q_i/\pss_i$, $y=q_j/\pss_j$, the terms inside the brackets in (\ref{gen-mid}) is expressed as
\eq{
f(y)-f(x)+(x-y)f'(y)=(x-y)\[ f'(y)-\frac{f(x)-f(y)}{x-y}\] ,
}
which is nonnegative due to the concavity of $f$.
This completes the proof of the relation (\ref{f-max}).


\section{Another example of the relation (3)}
The second example of the relation \eqref{relax} is a Brownian particle trapped by a harmonic potential in one dimension, which demonstrates the validity of the relation (3) for the spatially continuous systems. The time evolution of the probability distribution $p(x,t)$ is given by the following Fokker-Planck equation:
\eq{
\frac{\del}{\del t}p(x,t)=D \frac{\del}{\del x}\( \beta kx+\frac{\del}{\del x}\) p(x,t),
}
where $k$ is a spring constant and $D$ is a diffusion coefficient.
The origin of the coordinate is set to the center of the harmonic potential.
The initial distribution of the particle is set as a Gaussian distribution:
\eq{
p(x,0)=\frac{1}{\sqrt{4\pi a}}e^{-(x-b)^2/2a},
}
where $a$ and $b$ are respectively the variance and mean of the initial probability distribution at $t=0$.
The exact solution of the probability distribution at time $t$ is given by
\eq{
p(x,t)=\frac{1}{\sqrt{4\pi a_t}}e^{-(x-b_t)^2/2a_t}, \nt
}
where we defined $a_t:=(1-g(t)^2)/\beta k +ag(t)^2$ and $b_t:=bg(t)$ with $g(t):=e^{-D\beta kt}$.
Using the distribution, the difference between $\sigma_{[0,\tau]}$ and $D(\bsp(0)||\bsp(\tau)$ is calculated as
\balign{
&\sigma_{[0,\tau]}-D(\bsp(0)||\bsp(\tau)  =\frac{(1-g(\tau)^2)(a\beta k-1)^2}{2(g(\tau)^{-2}+a\beta k-1)}
+\frac{b^2\beta}{2}\frac{a\beta k(1-g(\tau)^2)+(2g(\tau)^{-1}-3+g(\tau)^2)}{g(\tau)^{-2}+a\beta k-1} ,
}
where we defined $g(\tau):=e^{-D\beta k\tau}$.
Owing to $g(\tau)\leq 1$, we find that the both denominators, the first numerator, and the first term in the second numerator are nonnegative.
In addition, the second term in the second numerator is also nonnegative, which is shown by applying the inequality of arithmetic and geometric means as $g(\tau)^{-1}+g(\tau)^{-1}+g(\tau)^2\geq 3\sqrt[3]{g(\tau)^{-1}\cdot g(\tau)^{-1}\cdot g(\tau)^{2}}=3$.
By combining them, our inequality \eqref{relax} in this setup is confirmed through a direct calculation of the probability distribution.

\figinb{7cm}{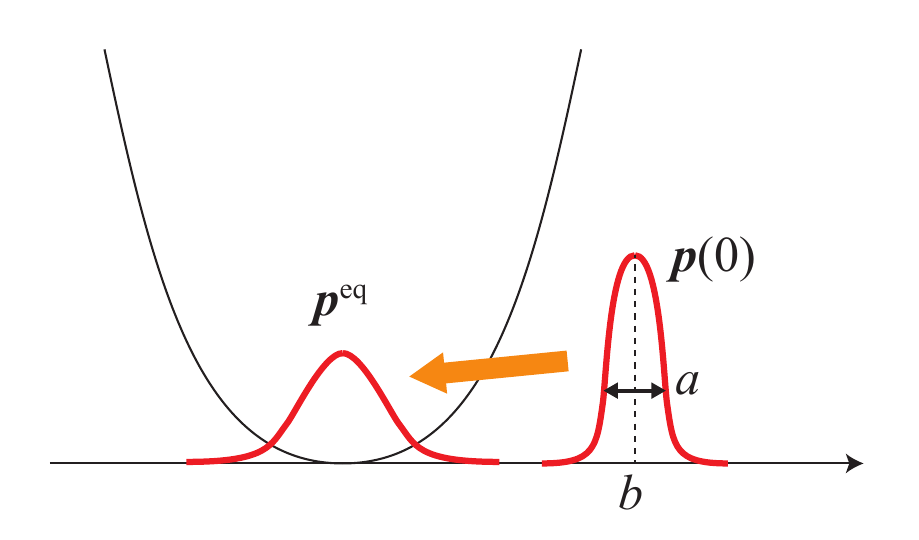} {Schematic of the second toy example for the relation (3) in the main text.}{harmo-ex}

\section{Tightness of (3) in local equilibration}

In this section, we show that if the relaxation process possesses time-scale separation and it realizes local equilibration, then the inequality \eqref{relax} achieves its equality. Without loss of generality, we suppose that states $1,\ldots, M$ concern fast relaxation, and states $M+1,\ldots, N$ concern slow relaxation.
In other words, the elements of the transition matrix $R_{ij}$ ($i$ or $j$ is in $\{ M+1,\ldots ,N\}$) is extremely small compared to the typical value of $R_{ij}$ ($i,j\in \{ 1,\ldots ,M\}$).
We assume that there exists a proper $\tau$ such that $p_i(\tau)$ ($1\leq i\leq M$) are in local equilibrium, while $p_i(\tau)=p_i(0)$ ($M+1\leq i\leq N$).
For later use, we define the probability in the states $1\leq i\leq M$ as $P:=\sum_{i=1}^M p_i(0)=\sum_{i=1}^M p_i(\tau)$.

Let us introduce an auxiliary system with $M$ states, whose transition matrix is given by $\bar{R}_{ij}:=R_{ij}$ ($i,j\in \{ 1,\ldots ,M\}$). We denote its probability distribution at $t$ and its equilibrium distribution by $\bar{p}_i(t)$ and $\bar{p}_i^{\rm eq}$, respectively. We denote the total entropy production between $0\leq t\leq \tau$ with the initial distribution is $\bar{p}_i(0)=p_i(0)/P$ by $\bar{\sigma}$.
We know the following expression:
\eq{
\bar{\sigma}=\sum_{i=1}^M \frac{p_i(0)}{P}\ln \frac{p_i(0)}{P\cdot \bar{p}_i^{\rm eq}}.
}

We now go back to the original system. Using the auxiliary system, the entropy production during the time interval $0\leq t\leq \tau$ is calculated as
\eq{
\sigma_{[0,\tau]}=P\bar{\sigma}.
}
At the same time, owing to $p_i(\tau)=P\bar{p}_i^{\rm eq}$, we have
\eq{
D(p(0)||p(\tau))=\sum_{i=1}^M p_i(0)\ln \frac{p_i(0)}{p_i(\tau)} =P\sum_{i=1}^M \frac{p_i(0)}{P}\ln \frac{p_i(0)}{P\cdot \bar{p}_i^{\rm eq}},
}
which confirms the tightness of the relation \eqref{relax} (the equality of the inequality \eqref{relax} is satisfied) at the local equilibrium.

\section{Derivation of the fluctuation-theorem type equality (\ref{FT-type})}

We here prove the fluctuation-theorem type equality \eqref{FT-type}:
\eq{
\la \exp \[ -\hat{\sigma}(\Gamma_{i\to j})-\ln \frac{p_j(\Di t)}{q_j(0)}+\ln \frac{p_i(0)}{q_i(\Di t)}\] \ra =1,
}
where $\hat{\sigma}(\Gamma_{i\to j})$ is the stochastic entropy production defined as $\hat{\sigma}(\Gamma_{i\to j}):=\ln [p_i(0)P(\Gamma_{i\to j})/p_j(\Di t)P(\Gamma^\dagger_{j\to i})]$~\cite{Sei12}. The quantity $P(\Gamma_{i\to j})$ is the probability for realizing the trajectory $\Gamma_{i\to j}$ under the condition that the initial and final states respectively are $i$ and $j$, and $\Gamma^\dagger_{j\to i}$ is the time reversal trajectory of $\Gamma_{i\to j}$.

The fluctuation-theorem-type equality is derived as follows:
\balign{
&\sum_{i,j} \int d\Gamma_{i\to j}p_i(0)P(\Gamma_{i\to j}) e^{-\hat{\sigma}(\Gamma_{i\to j})-\ln \frac{p_j(\Di t)}{q_j(0)}+\ln \frac{p_i(0)}{q_i(\Di t)}} \nt \\
=&\sum_{i,j} \int d\Gamma^\dagger_{j\to i}\frac{p_i(0)}{q_i(\Di t)}q_j(0)P(\Gamma^\dagger_{j\to i}) \nt \\
=&\sum_{i} \frac{p_i(0)}{q_i(\Di t)}q_i(\Di t) =1.
}
In the derivation of the conventional fluctuation theorem, there exists arbitrariness in the choice of the initial distribution of the backward path.
Our derivation of \eref{FT-type} employs this fact.
We also note that the forward transition probability $P(\Gamma_{i\to j})$ and the backward transition probability $P(\Gamma^\dagger_{i\to j})$ are the same, which results from the time-independence of transition rates and the detailed balance condition.
Owing to this fact, the transformation in the third line, $\sum_j \int  d\Gamma^\dagger_{j\to i}q_j(0)P(\Gamma^\dagger_{j\to i})=\sum_j \int  d\Gamma_{j\to i}q_j(0)P(\Gamma_{j\to i})=q_i(\Di t)$, is justified.

For the general case where the transition rates are time-dependent and $\Delta t$ is a finite time duration, by imposing the dynamics ${\bm R} (t)={\bm R}(\Di t-t)$ for $t\in [0,\Delta t]$ in the above derivation, the same fluctuation-theorem-type equality (\ref{FT-type}) is readily obtained.

\end{widetext}

\end{document}